# Large individual differences in free recall

Eugen Tarnow

18-11 Radburn Road

Fair Lawn, NJ 07410

etarnow@avabiz.com

## Abstract

Using single factor ANOVA I show that there are large individual differences in free recall ($\eta^2$ ranges from 0.09-0.26) including the total recall, the balance between recency and primacy, and the initial recall (subsequent recalls show smaller individual differences).

All three memory properties are relatively uncorrelated. The variance in the initial position may be a measure of executive control and is correlated with total recall (the smaller the variation, the larger the recall).

## Introduction

Free recall stands out as one of the great unsolved mysteries of modern psychology REVIEWS. Words in a list are displayed or read to subjects who are then asked to retrieve the words. It is one of the simplest ways to probe short term memory but the results (Murdock, 1960; Murdock, 1962; Murdock, 1975) have defied explanation. Why do we remember primarily items in the beginning and in the end of the list, but not items in the middle, creating the famous u-shaped curve of probability of recall versus serial position? Why can we remember 50-100 items in cued recall or recognition but only 6-8 items in free recall?

While free recall has been used to diagnose memory impairments (Bayley et al, 2000; Buschke, 1999; Bemelmans, 2009; Tarnow, 2012) and as a test of working memory capacity (Watkins, 1974; Unsworth, 2007) there is only recently that memory psychology has focused on individual differences.

Unsworth, Spillers & Brewer (2010) shows that IF there are individual differences, these differences vary separately for primary and secondary memory, where the two memory stores are defined by the Tulving & Colotla's (1970) procedure. (Tulving & Colotla (1970) defined the capacity of primary memory as the sum of probabilities of recalling items with "intratrial retention interval" (the number of other items presented or recalled between the item presentation and recall) of 7 or less (Tulving & Colotla, 1970). This non *ab-initio* approach sets an *a priori* limit on the intratrial retention interval, thereby indirectly fixing the capacity of working memory. It also, presumably incorrectly, disallows first item recalls, discounts early recalls which include both the first and last items, and discounts recalls of the recency items depending upon the order of those recalls. Watkins (1974) obtains a primary memory capacity using this method which ranges from 2.93-3.35 words, presumably including cancellations of systematic errors. The model is not *ab-initio* with respect to either capacity nor the item content of working memory. However, if there is any partial truth to their conjecture, the procedure of Unsworth, Spillers & Brewer (2010) should work. That it does is shown by their self-consistent factor analysis of a 10 item free recall indicating that one factor includes the first 7 items and the other factor includes the five last items as well as the first item.)

Unsworth, Brewer and Spillers (2011) used an SPSS clustering algorithm on free recall subjects and identified three clusters: those with more primacy, those with more recency and those with both. The quality of the cluster identification was not reported. ANOVA results reported include a significant interaction of cluster groups and the serial position curve, cluster groups and probability of recalling the first three and last three items, and cluster groups and probability of recalling the first three and last three items in the first recall.

Healy & Kahana (2014) expressed skepticism about cluster analysis suggesting most of the algorithms rely on arbitrary rules of thumb or generic corrections for overfitting. Indeed, these authors also searched for individual differences in the serial position curve and were not able to reject the null hypothesis of a single cluster, in contradiction to the Unsworth, Brewer and Spillers (2011) result. They commented “However, the participants completed only 10 free recall lists (compared with 112 lists for the participants reported here), and therefore the subgroups may represent variation in strategy exploration and parameter tuning and not variation in core processes.” Thus Healy & Kahana (2014) is suggesting that to truly understand free recall data, ten lists is not enough but 112 lists might be.

I note that clustering in the number of dimensions at hand (ten dimensions for the ten free recall items) is extremely difficult to visualize. I also note that if there are a few well defined clusters, they would likely point to a single gene being responsible for the serial position curve which would be such an enormously important finding that there had to be scientific consensus as this clustering.

Healey, Crutchley and Kahana (2014) used a complex computer algorithm to identify four sources of variation IF there were individual differences. These sources were primacy, recency, temporal and semantic.

In this brief article I want to settle whether there are individual differences in free recall with a standard ANOVA analysis, whether there are clusters will have to be deferred to another article. I find that there are individual differences to be found in free recall including the total recall, the balance between recency and primacy, and the initial recall. My analysis .complements the efforts of Unsworth, Spillers & Brewer (2010), Unsworth, Brewer and Spillers (2011), Healey, Crutchley and Kahana (2014) and Healey and Kahana (2014).

## Method

This article makes use of the Murdock (1962) data set (downloaded from the Computational Memory Lab at the University of Pennsylvania (http://memory.psych.upenn.edu/DataArchive). In Table 1 is summarized the experimental processes which generated the data sets used in this paper. In this paper we focus on the smallest and largest tests – 10-2 and 40-1. The 10-2 test presented 10 items verbally at 2 second intervals and the 40-1 test presented 40 items verbally at single second intervals.

## Results

Fig. 1 displays the distribution of total recall. There are individual differences as shown by a single factor ANOVA for the 10-2 (upper panel) and 40-1 series (lower panel) with F=30 and F=18, respectively both with the critical value of F at 1.7 for alpha=0.05 (see Table 2. In this investigation alpha=0.05 for all calculations).

In Fig. 2, top panel, is shown the probability of recall by item averaged over all 15 subjects of the 10-2 series of Murdock (1962). The recall probability curve is fitted with a straight line using linear regression. I define the slope of this line as the recency primacy balance (RPB). A positive RPB indicates preference for recency and a negative slope indicates a preference for primacy. The average RPB is positive, 0.036. The same results grouped in halves with low (high) RPB is displayed in the middle (bottom) panel of Fig. 2.

In Fig. 3 is shown the distribution of RPB for the 10-2 series (top panel) and the 40-1 series (bottom panel). A single factor ANOVA results in F=19 (F=14) for the 10-2 (40-1) series.

Are total recall and RPB related? In Fig. 4 is displayed individual averages of the total recall and RPB together with a linear regression. The measures are not strongly correlated. For the 10-2 series (upper panel) Rsquared=0.3 and for the 40-1 series (bottom panel) Rsquared=0.007.

Effect sizes are calculated in Table 2. They are the largest for the 10-2 data and the largest for total recall. In five of the six cases the effect sizes are large using Cohen's (1988) guidelines.

Fig. 5 displays the F values for increasing number of trials. Total recall F-1 is proportional to the total number of trials. In contrast, RPB starts out with below-critical F values for the 40-1 series which increase linearly beyond critical at about the twentieth trial. The 10-2 series starts out with above critical F values which increase linearly also at about the twentieth trial. In both cases the between-group variation is responsible for the non-linearity.

Do averages change with the number of trials? Fig. 7 displays how the total recall and RPB varies with trial. The variation is a small decrease in total recall and no change at all in RPB.

Other individual differences include the initial recall. In order to perform an ANOVA analysis the underlying distribution needs to be somewhat normal (ANOVA is relatively insensitive to changes to the normal distribution). For the first recall this is clearly not the case – there are two peaks – the first item and the last items. This is a simple numerical issue: I rename the first items 1 … N/2 as N+1 … N+N/2. In Fig. 8 is displayed the first item recall on such a scale and it shows a single peak. A single factor ANOVA gives F=20 (F=8.5) for the 10-2 and 40-1 series. In both cases F-1 is proportional to the number of trials (Fig. 9).

The average initial position is relatively uncorrelated with total recall (see Fig. 10) and with RPB (see Fig. 11).

Could the variance in initial position be related to executive control? If so, does a low variation correlate with total recall? In Fig. 12 is shown that individual differences in the variance of the initial position are weakly correlated with the total recall, small variations are associated with high recalls.

Recalls after the initial recall show less pronounced individual differences as displayed in Fig. 13.

## Discussion

Using ANOVA I have shown that there are individual differences in free recall including total recall, initial recall and the balance between recency and primacy. The three properties have large effect sizes (largest for the 10-2 data) and are relatively uncorrelated. For the 40-1 data, the balance between recency and primacy emerges as statistically significant only after some twenty trials (the same pattern occurs in the 10-2 series though the initial F is larger than Fcritical due to a single individual with a large negative recency primacy balance).

These results have similarities with the ones of Unsworth, Brewer and Spillers (2011); the participants in the three clusters differed in their tendency for primacy and recency overall and the primacy and recency of the first recall as well as the overall serial position curve. As mentioned, if there truly are three clusters of free recall participants it would be extremely important, presumably indicating that one or two genes are responsible for the serial position curve in free recall. Healy & Kahana (2014) did not find a cluster in the serial position curve and proposed that individual differences come from quantitative variations rather than qualitative variations. In my analysis I do not see multiple distributions, just wide distributions (Figs. 1,3 and 8), but the data I analyzed had only 15 participants per experiment, in the data of Unsworth, Brewer and Spillers (2011) there were many more, 150 participants.

Large effects in free recall confirms that free recall is a good component of aptitude measures.

Significant individual differences in the balance of recency and primacy present a mystery. They are not related to total recall or initial recall. The balance between recency and primacy is of some importance: it is a sensitive measure of the presence of several diseases such as Alzheimer's disease (primacy disappears for Alzheimer's disease, see Bayley et al, 2000, Buschke, 1999 and Tarnow, 2012) and of the presence of drugs such as marijuana (Tarnow, 2012).

There is no literature as to what the initial recall might suggest about a person's memory function, though the variance might suggest a measure of executive control. If the variance is low, the subject might be controlling the initial choice. The effect of executive control is a maximal 0.7 items for the 10-2 data and 2 items for the 40-1 data.

# TABLES

| *Work* | *Item types* | *List length and presentation interval* | *Recall interval* | *Subjects* | *Item presentation mode* |
|---|---|---|---|---|---|
| *Murdock (1962)* | *Selection from 4000 most common English words, referred to as the Toronto Word Pool.* | *40 words in a list, each word presented once a second*<br><br>*10 words in a list, each word presented once every two seconds* | 1.5 minutes | 103 undergraduates divided between six experiments. | Verbal. 80 trials. |

*Table 1. Experimental method that generated the data used in this contribution.*

| | *10-2 total recall* | *40-1 total recall* | *10-2 RPB* | *40-1 RPB* | *10-2 initial recall* | *40-1 initial recall* |
|---|---|---|---|---|---|---|
| *F* | *29.7* | *17.6* | *19.1* | *13.8* | *20.3* | *8.37* |
| *P* | *9E-68* | *5E-40* | *9E-44* | *7E-31* | *3E-46* | *2E-17* |
| *Effect size ($\eta^2$)* | *0.26 (large)* | *0.17 (large)* | *0.18 (large)* | *0.14 (large)* | *0.20 (large)* | *0.094 (medium-large)* |

*Table 2. ANOVA results showing individual differences. Alpha=0.05 for all calculations. The effect sizes are scored as “large” or “medium-large” using Cohen’s (1988) guidelines.*

FIGURES

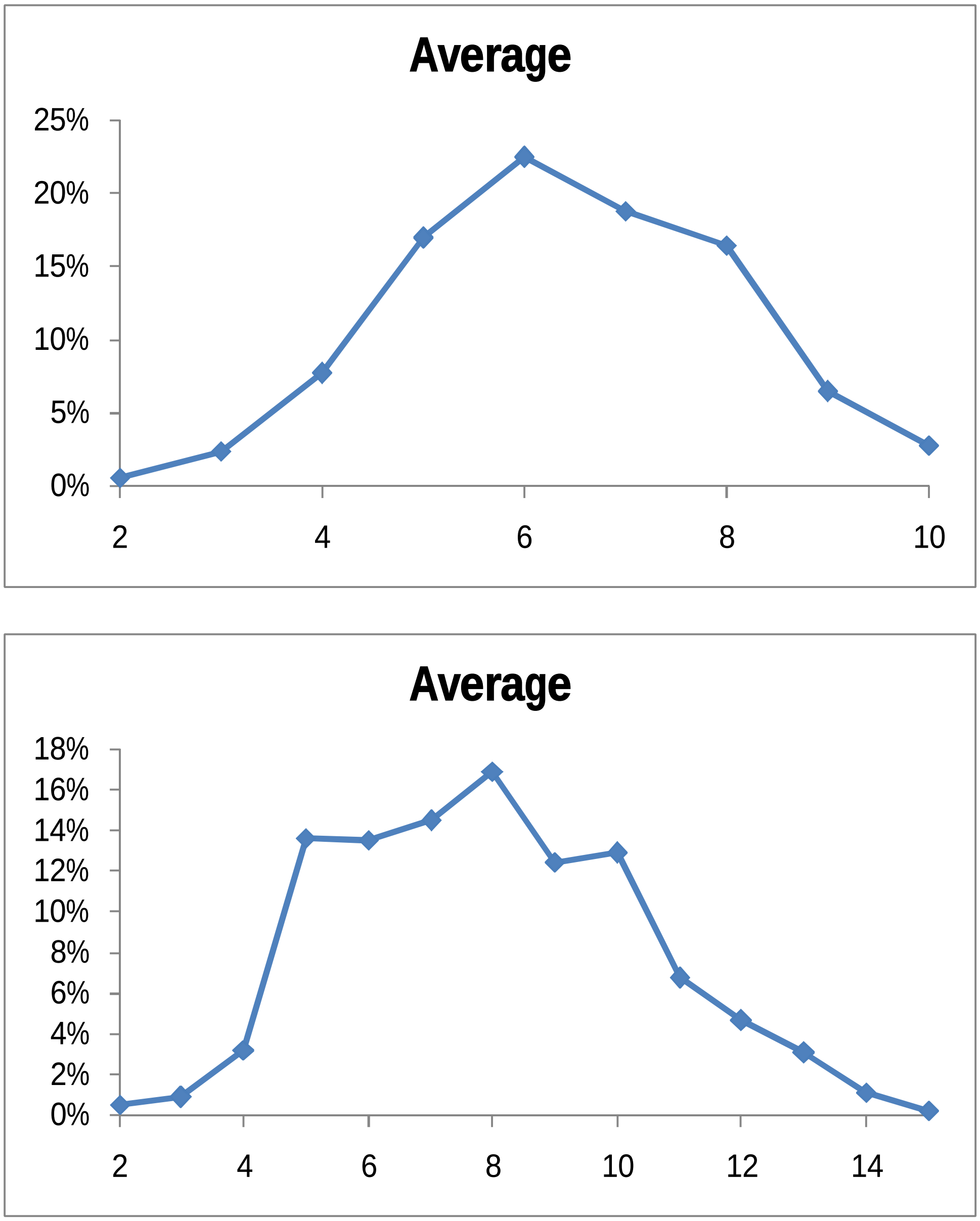

Fig. 1. Distribution of total recall. Upper panel results from the 10-2 series (single factor ANOVA yields F=29.7 and p 9ee-68; lower panel results from the 40-1 series (F=17.6, p=5e-40).

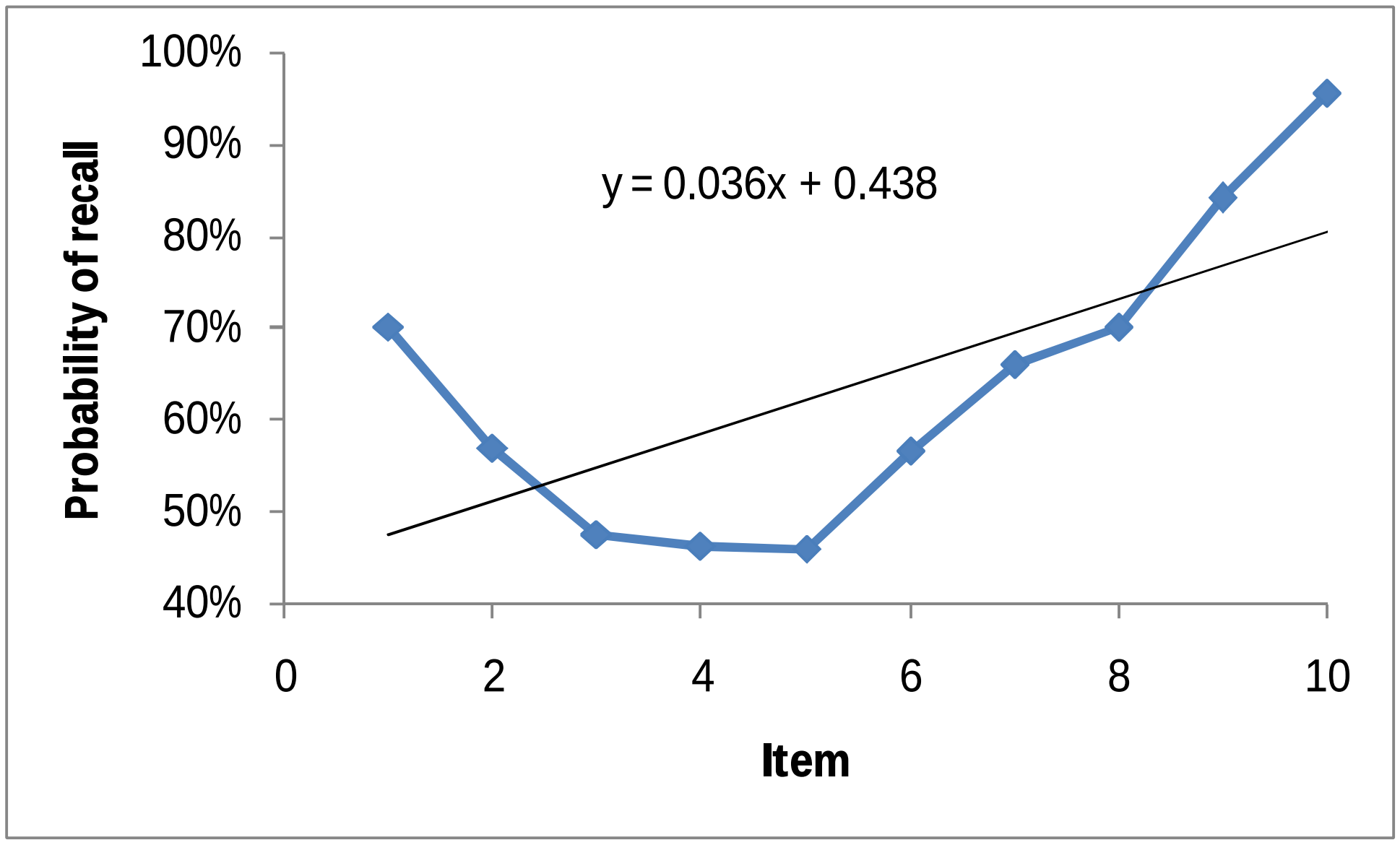

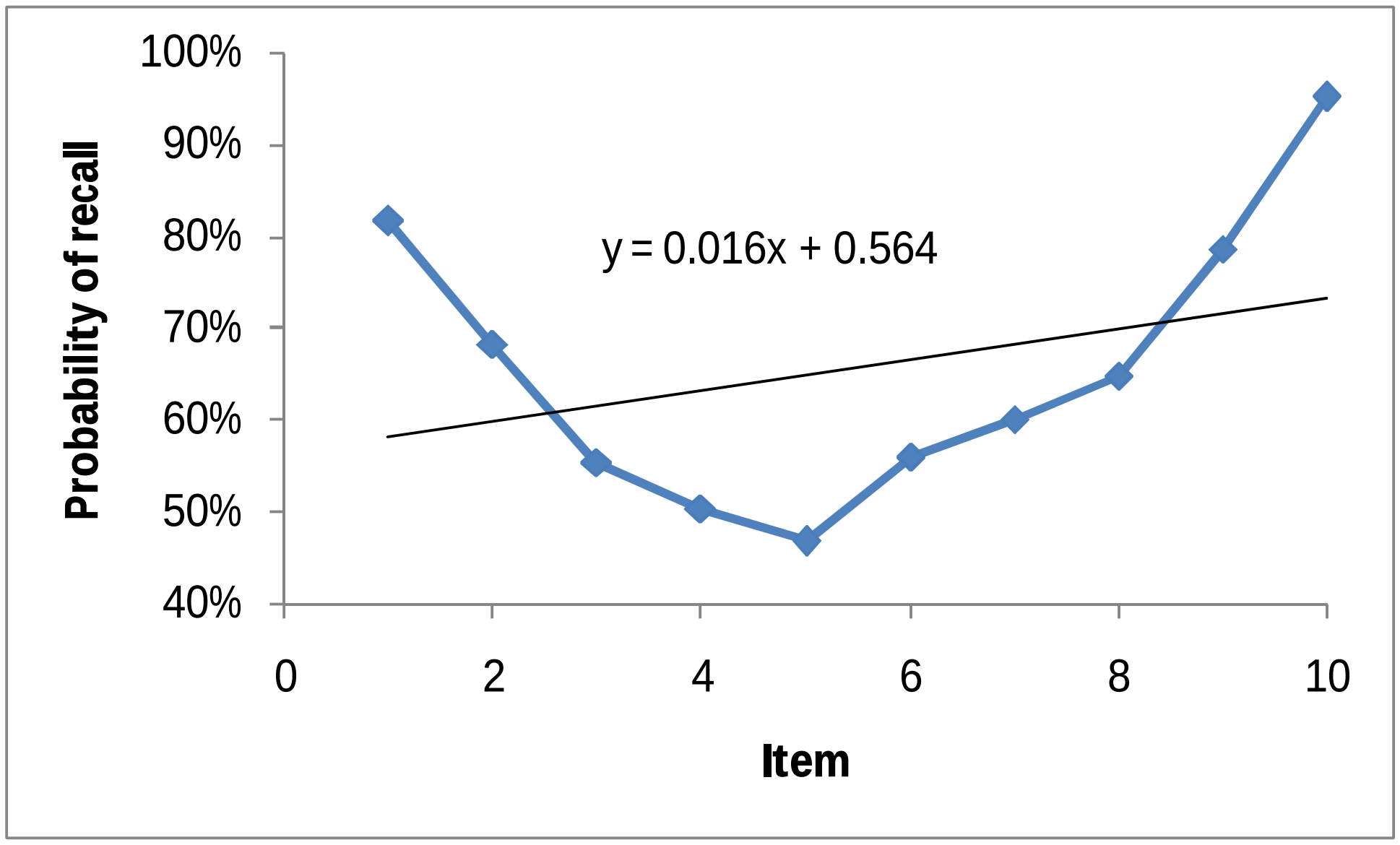

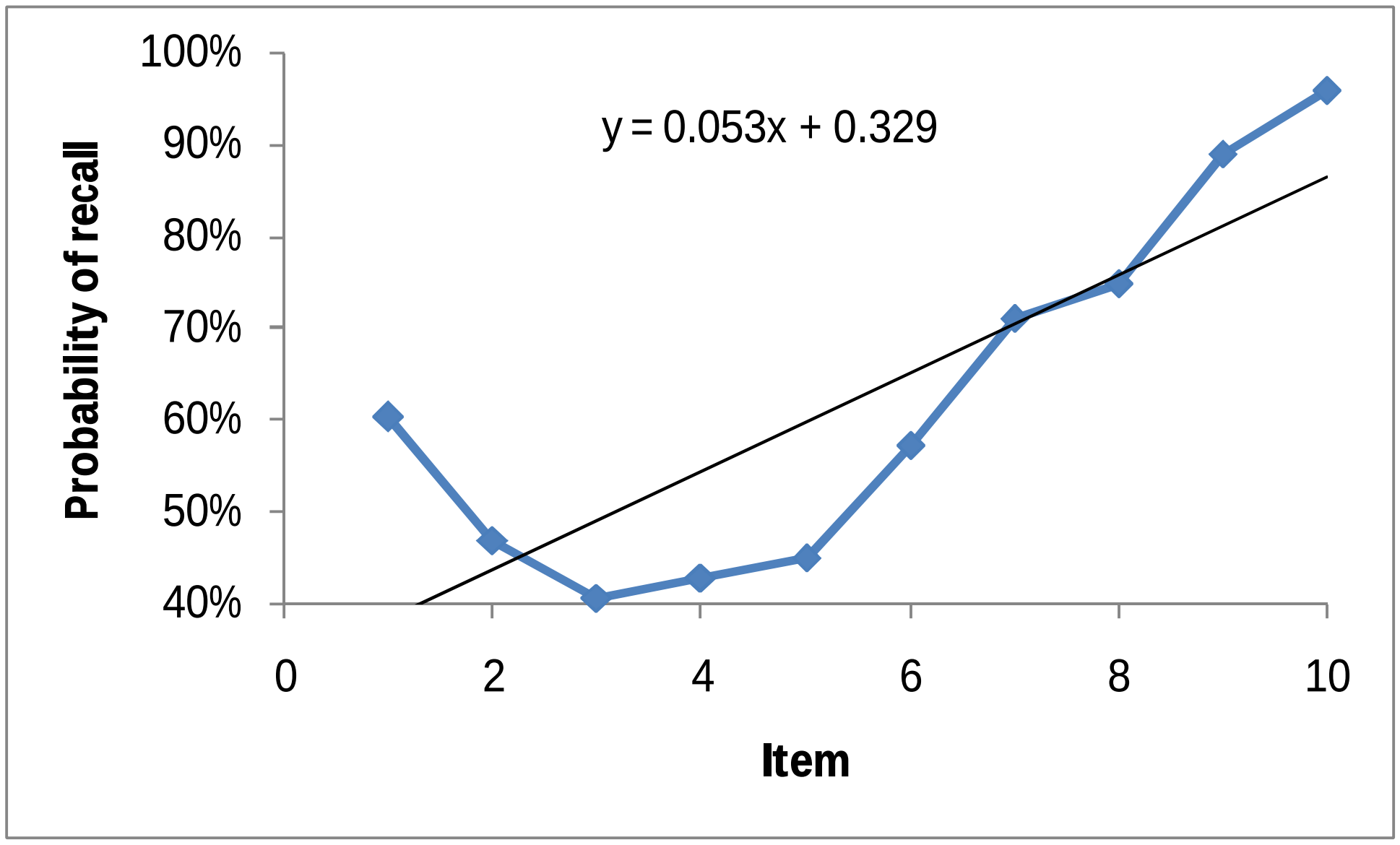

Fig. 2. Probability of recall for 10-2 series (top panel). T he slope of the straight line regression is defined as the recency primacy balance (RPB). Middle panel shows results for individuals with low recency primacy balance and bottom panel shows results for individuals with high recency primacy balance.

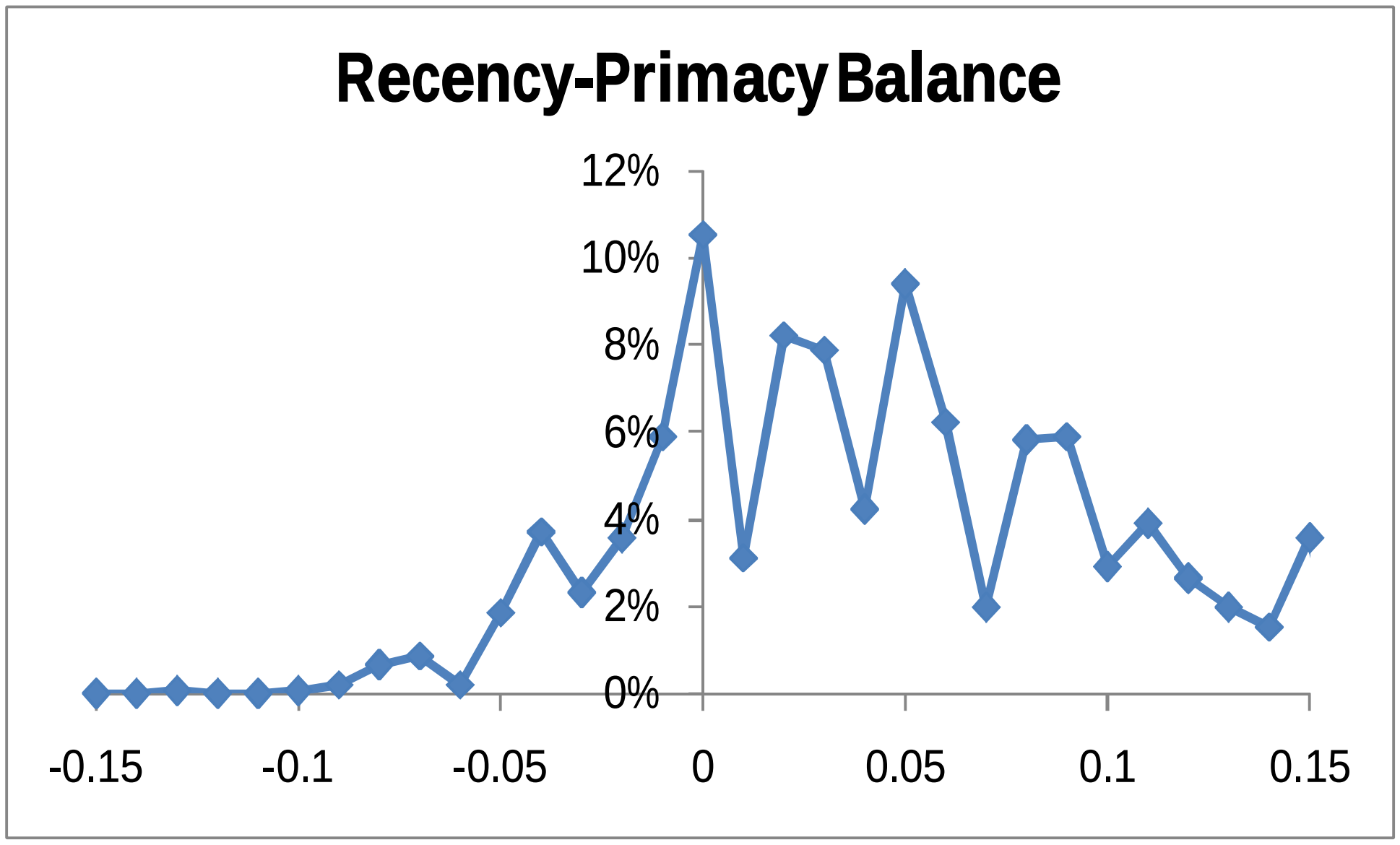

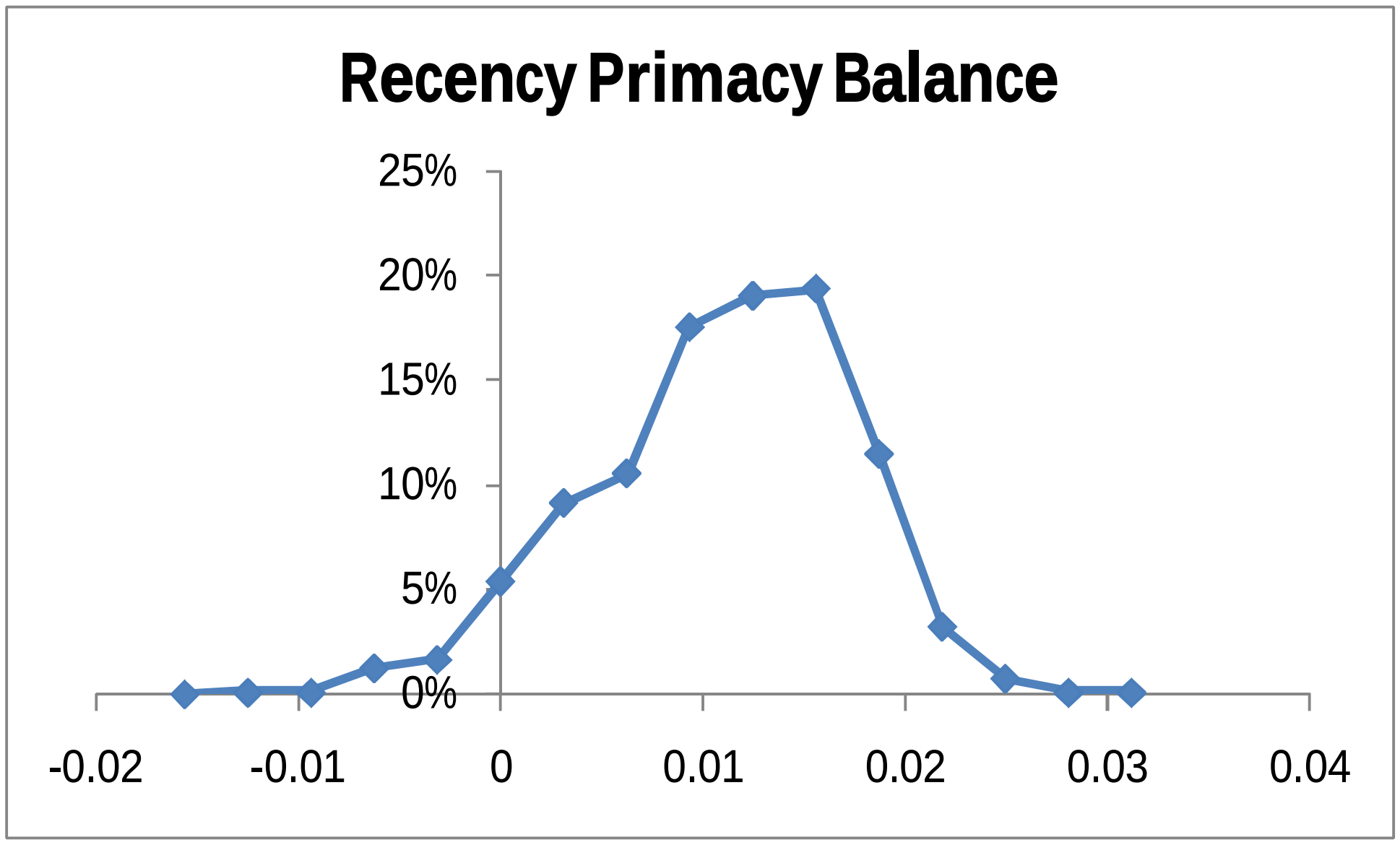

Fig. 3. Recency primacy balance for 10-2 series (top panel) and 40-1 series (bottom panel).

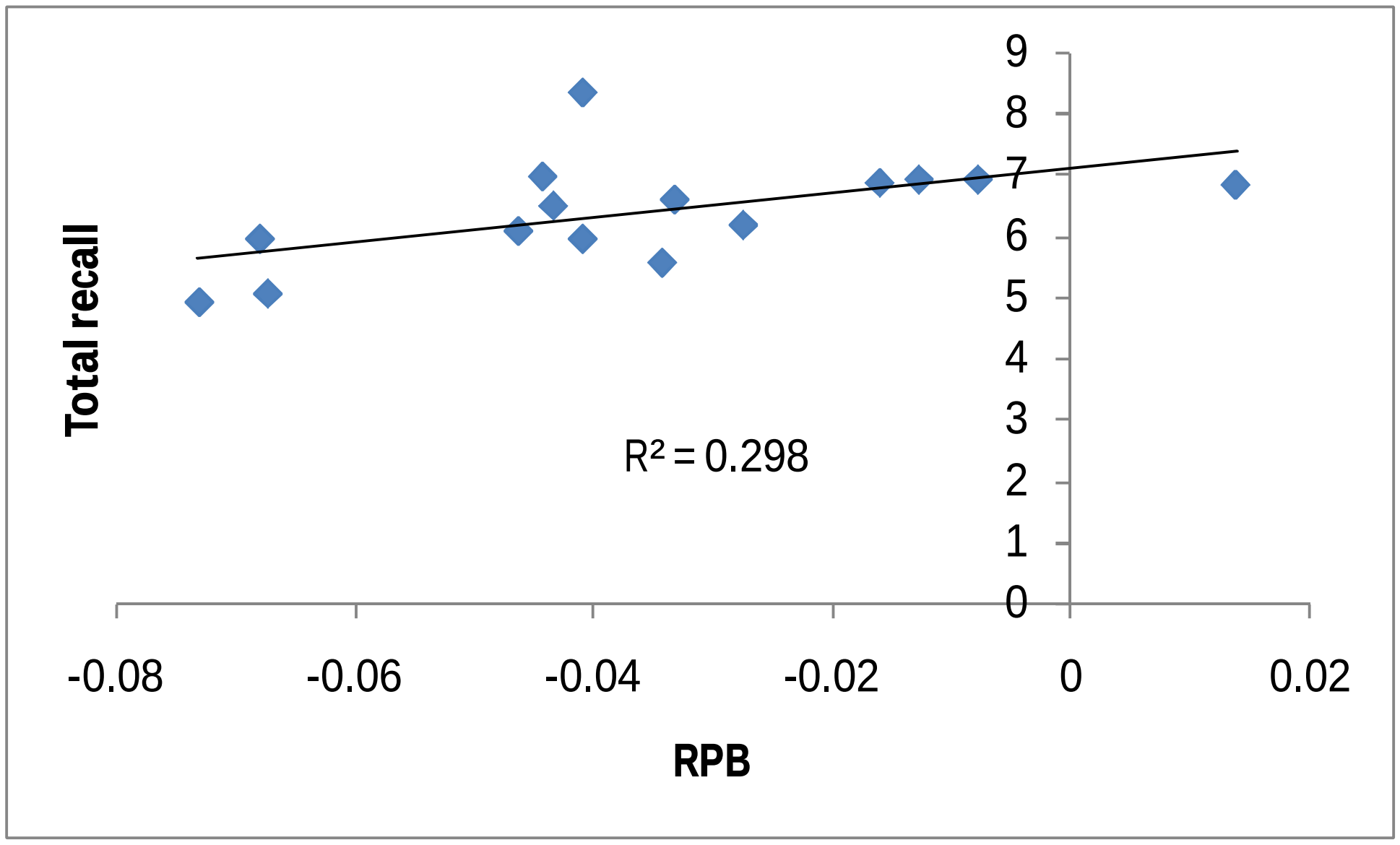

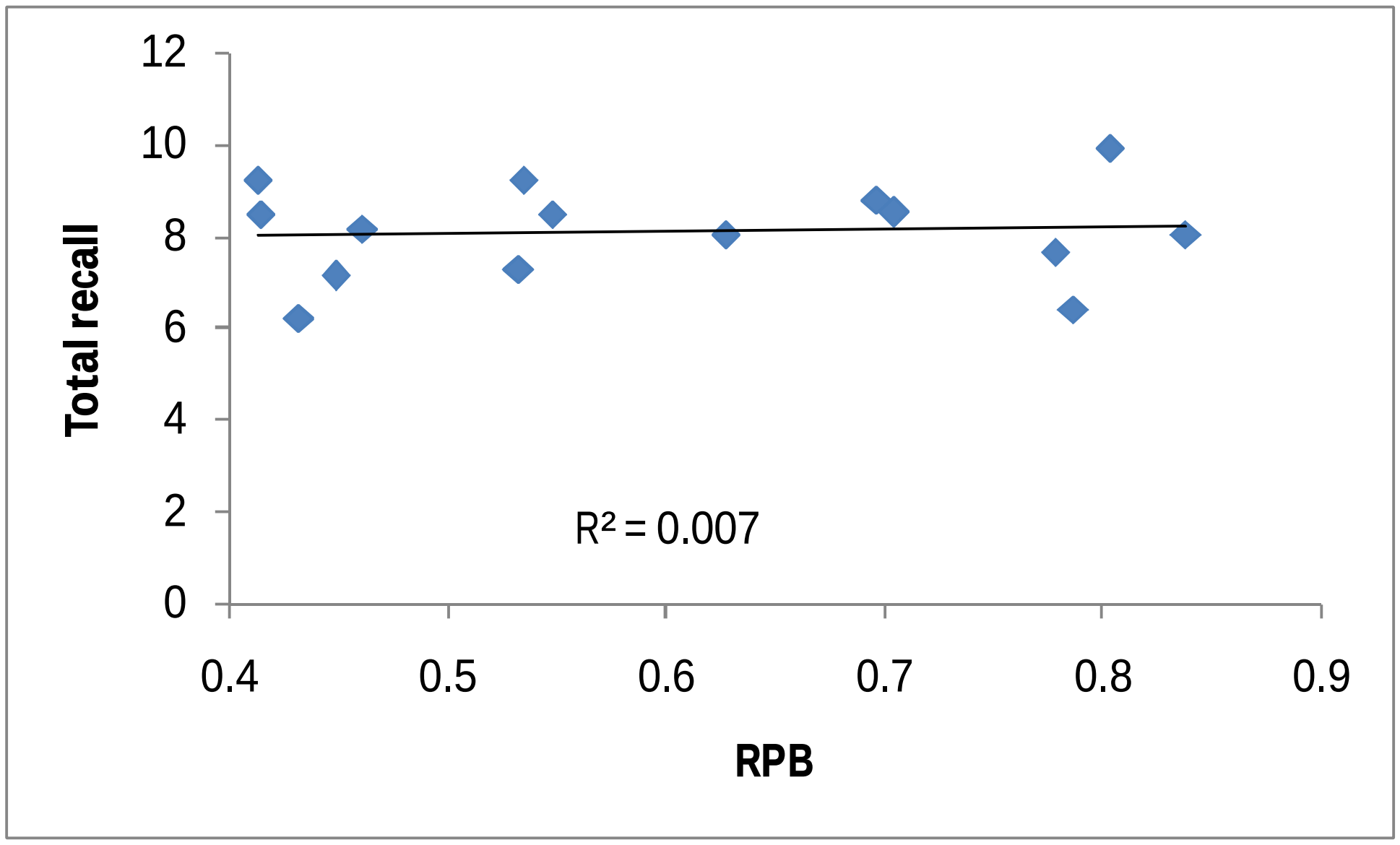

Fig. 4. Individual scores of total recall and RPB together with a linear regression line. Upper panel displays 10-2 data and the lower pandel displays 40-1 data.

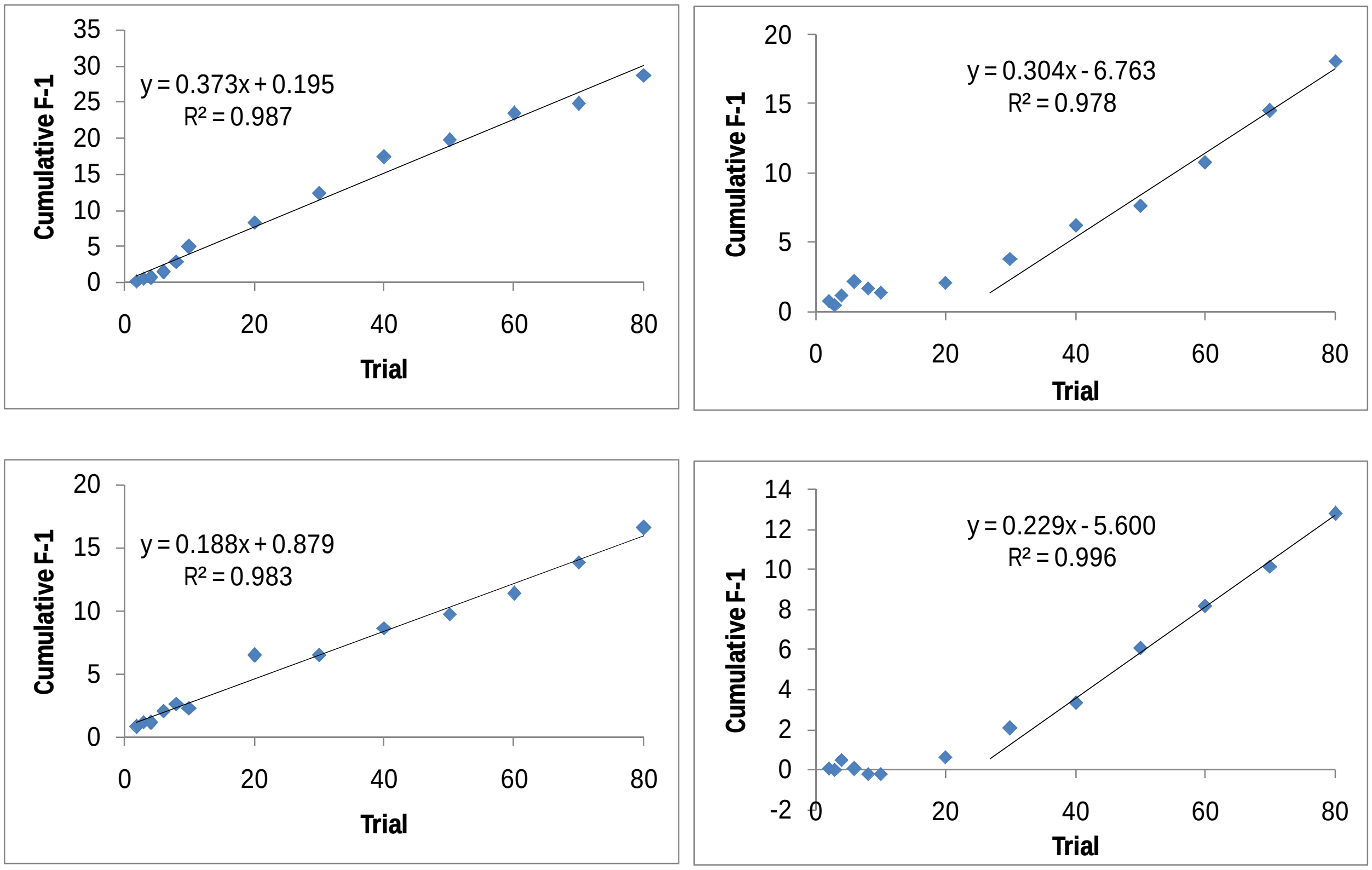

Fig. 5. Cumulative F values for the 10-2 data (upper panels) and 40-1 data (lower panels). Left panel displays the total recall and the right panel displays RPB. Notice the deviation from proportionality for the RPB data.

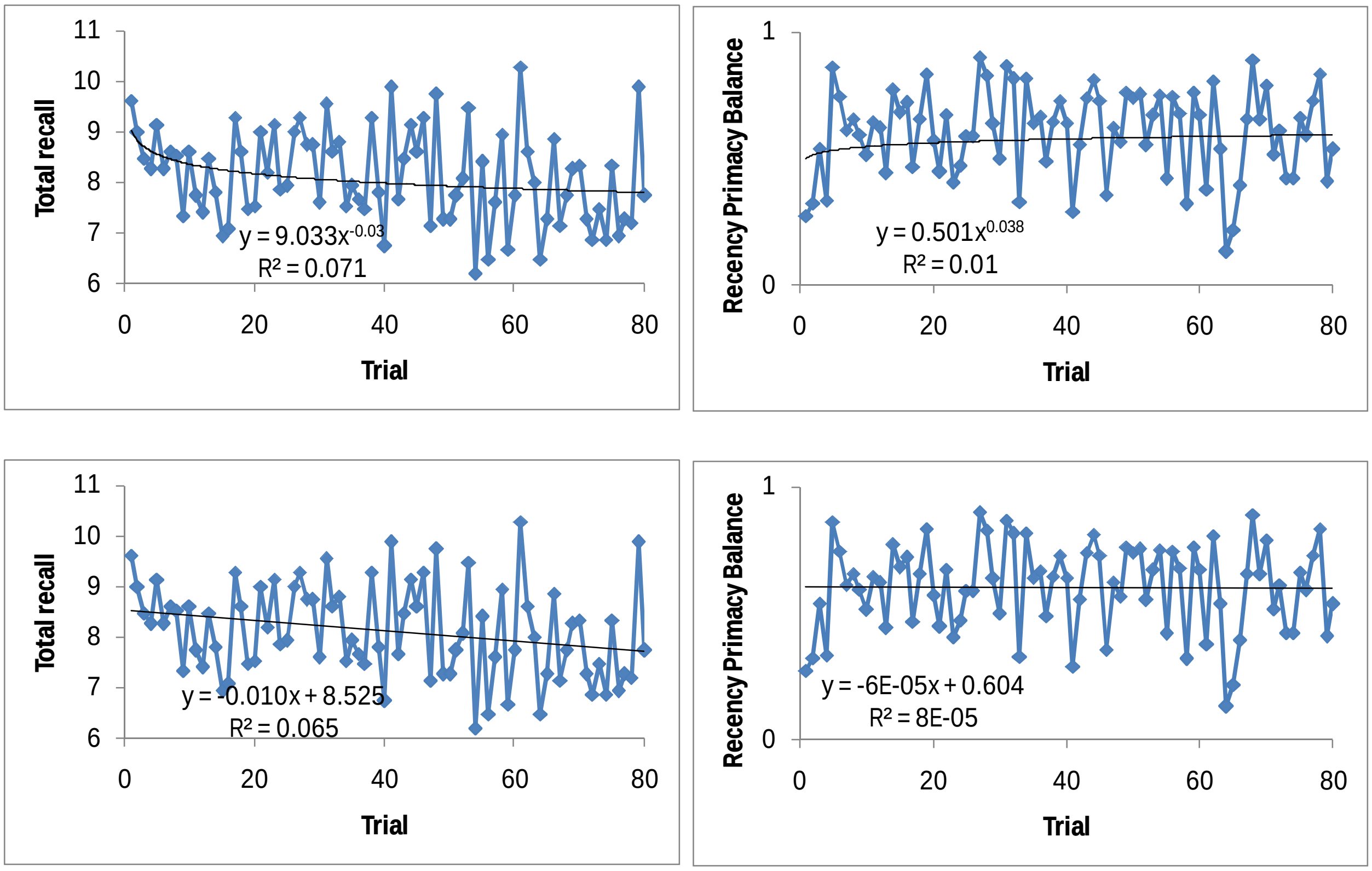

Fig. 7. Total recall (left panels) and recency primacy balance (right panels) as a function of trial for 40-1 data- power fit (top panels) and linear fit (bottom panels). Note that neither measure changes much with trial.

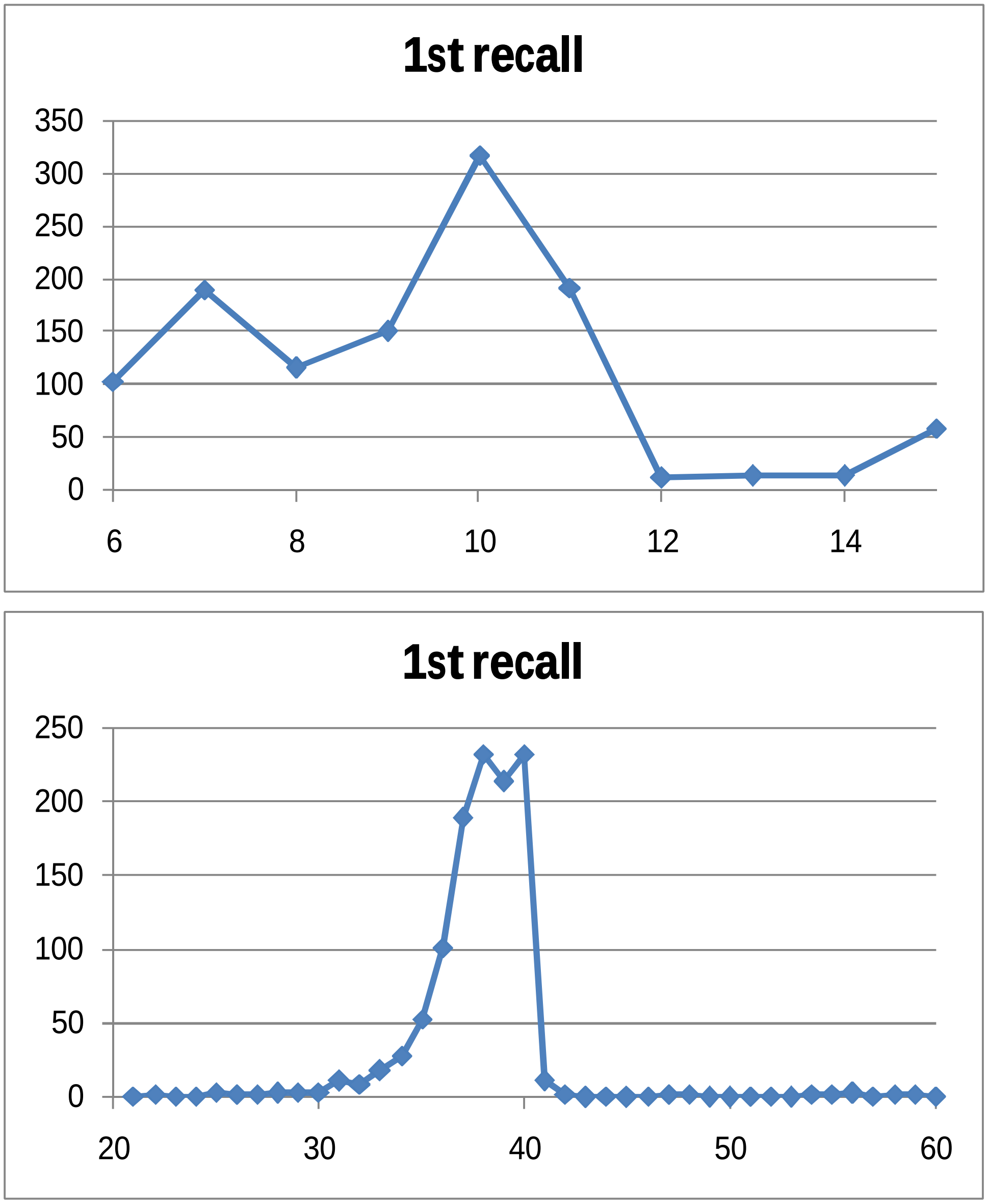

Fig. 8. 1st recall distribution for the 10-2 (top panel) and 40-1 series (bottom panel).

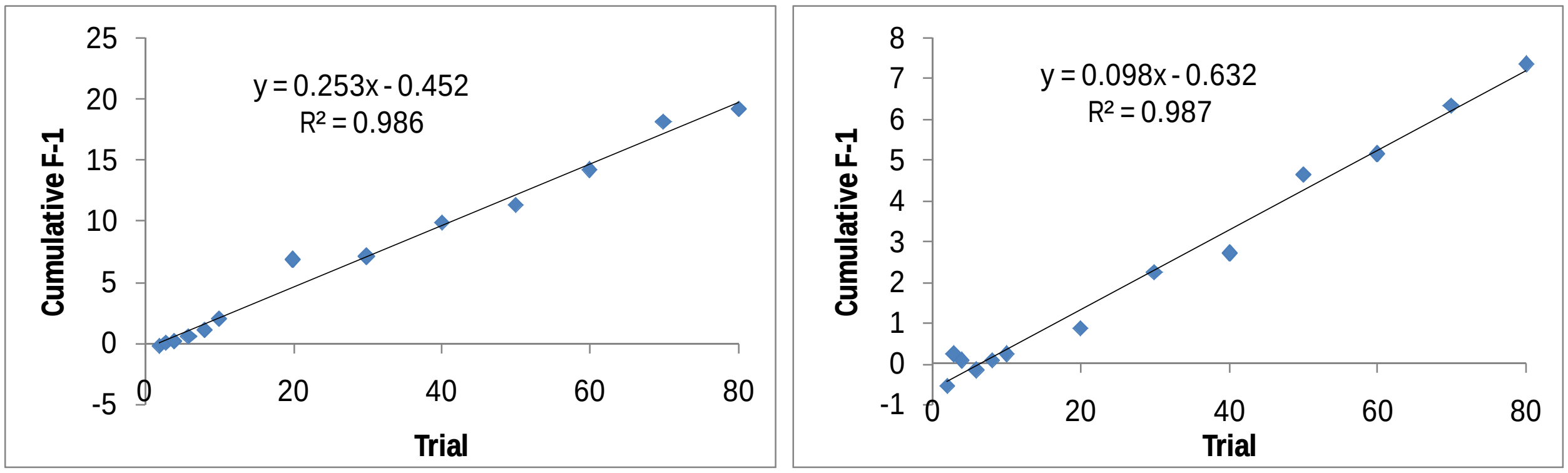

Fig. 9. F as a function of trial for the initial recall. 10-2 data on left, 40-1 data on right.

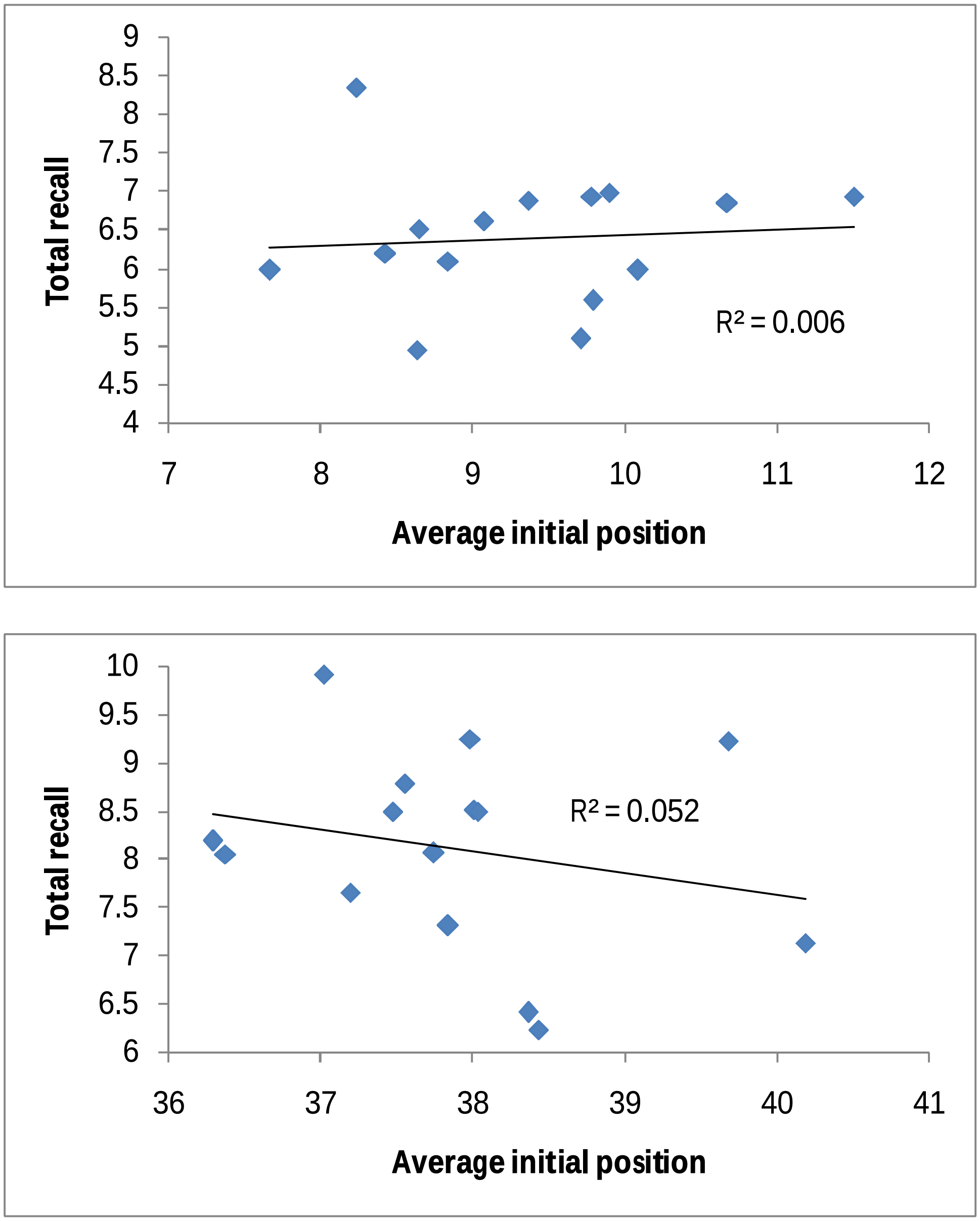

Fig. 10. Individual total recall as a function of the average initial positions for the 10-2 series (upper panel) and the 40-1 series (lower panel). See text for the meaning of initial positions larger than 10 and 40, respectively.

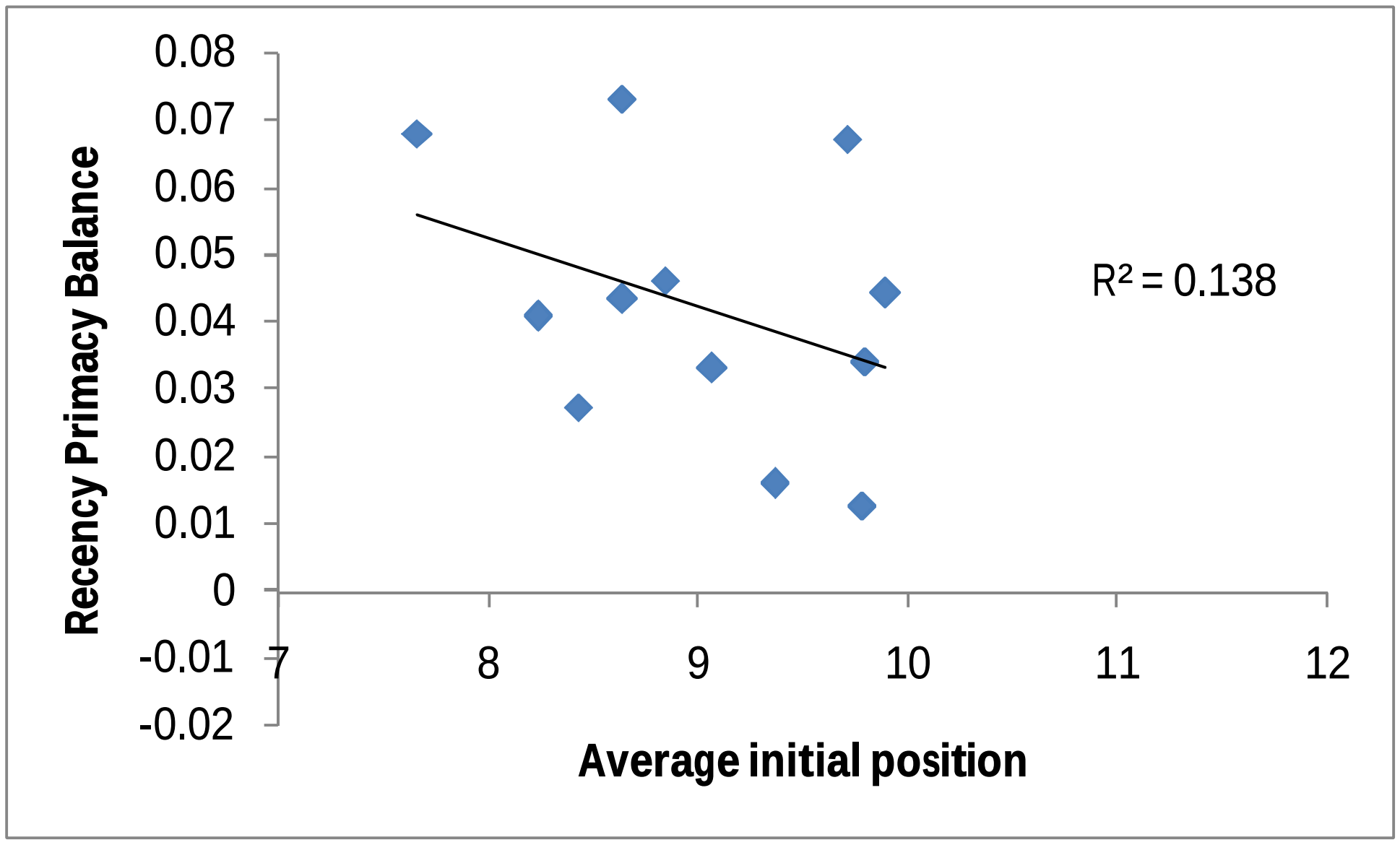

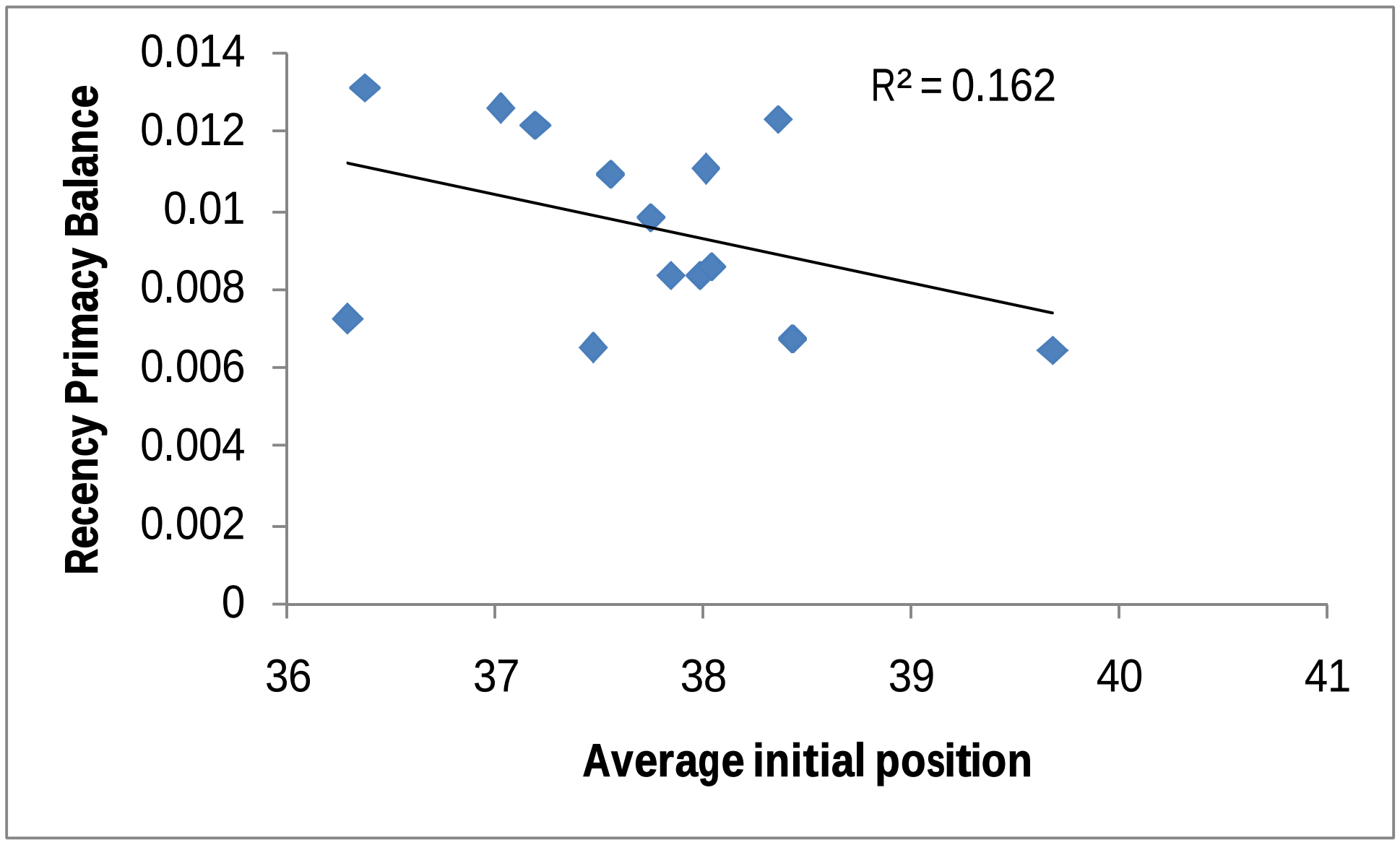

Fig. 11. RPB versus average initial positions for the 15 individuals of the 10-2 series (upper) and 40-1 series (lower panel) for which the average initial positions were less than 40 and 10, respectively.

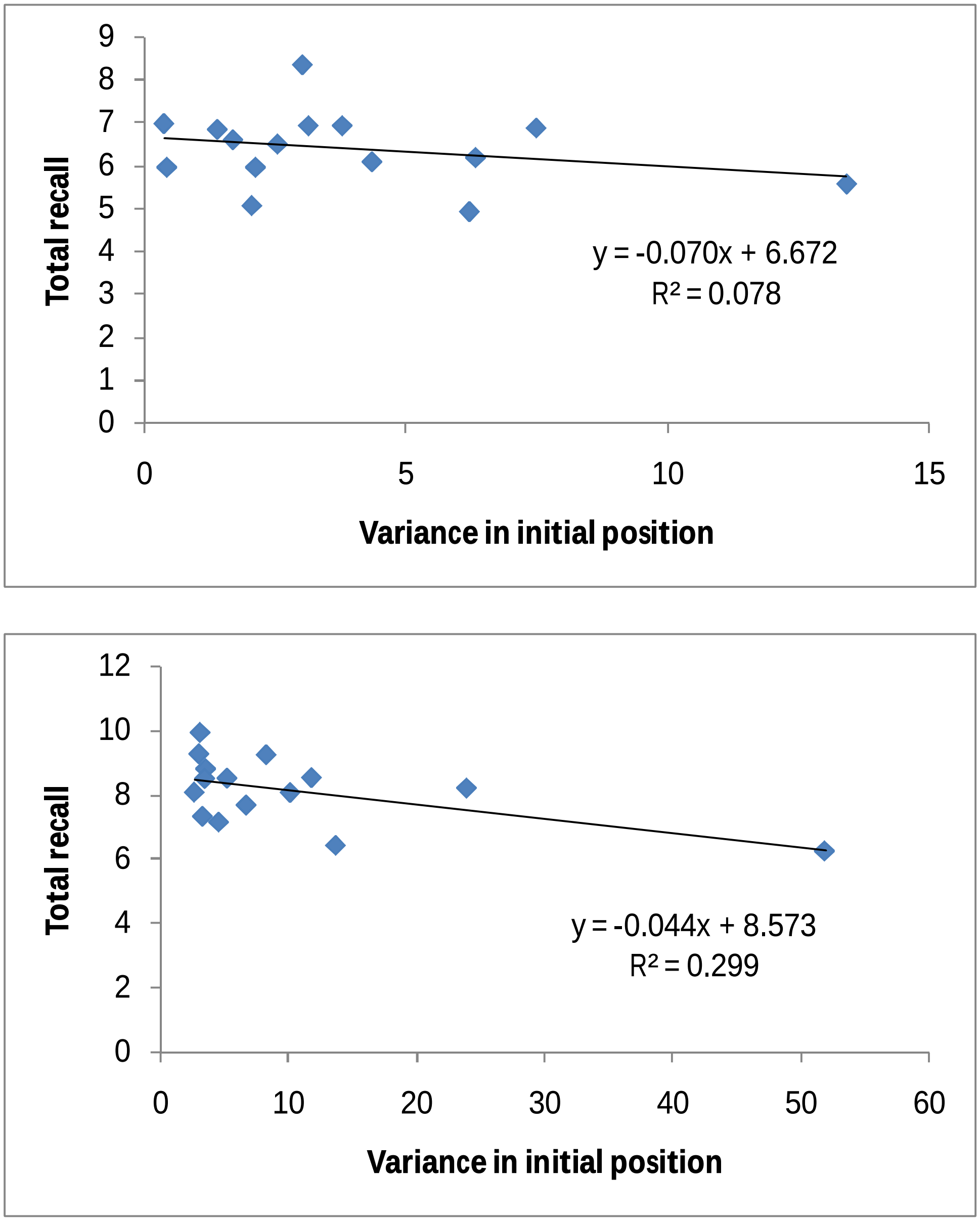

Fig. 12. Correlation between variance in initial position and total recall for the 10-2 series (top panel) and the 40-1 series (bottom panel). A very low variance (high control) increases the total recall by at most 0.7 items for the 10-2 series and at most 2 items for the 40-1 series.

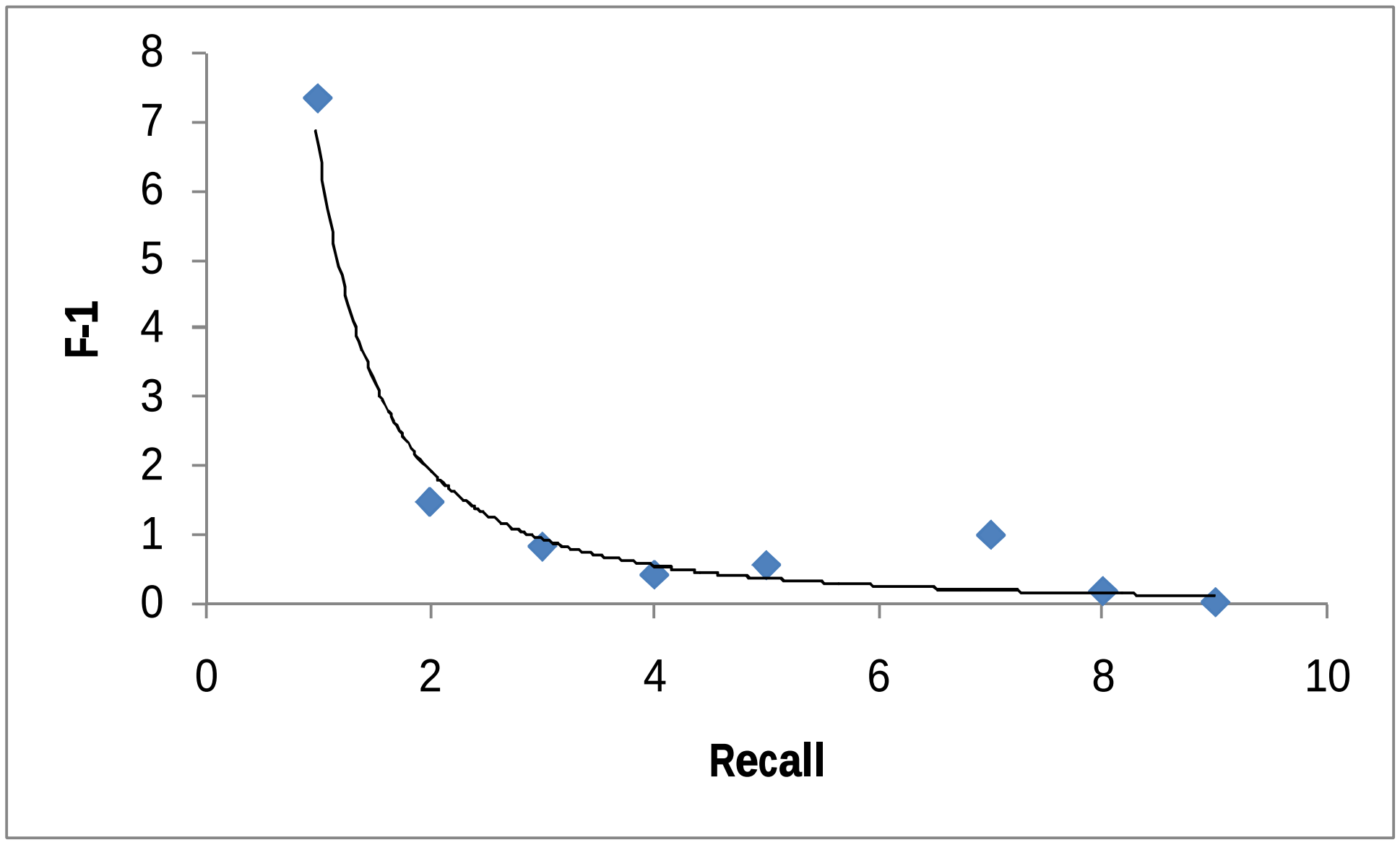

Fig. 13. F values as a function of the recall. The value for the sixth recall was -0.11 and not included in the curve fit. The critical value of F was 1.7, so the critical value of F-1 was 0.7.